\documentclass[preprint,showpacs,preprintnumbers,amsmath,amssymb,prb]{revtex4}

\usepackage{graphicx}% Include figure files
\usepackage{dcolumn}% Align table columns on decimal point
\usepackage{bm}% bold math

\begin{document}

%\preprint{APS/123-QED}

\title{Electron Spin Relaxation and $^{39}$K Pulsed ENDOR Studies on Cr$^{5+}$ doped K$_3$NbO$_8$ at 9.7 and 240 GHz.}

\author{S. Nellutla}
\affiliation{Center for Interdisciplinary Magnetic Resonance,
National High Magnetic Field Laboratory, Florida State University,
Tallahassee, Florida-32310, USA.}
\author{G. W. Morley}
\altaffiliation[Current address: ]{London Center for Nanotechnology,
University College London, London-WC1H0AH, UK.}
\author{J. van Tol}
\email[Corresponding author. ]{vantol@magnet.fsu.edu}
\affiliation{Center for Interdisciplinary Magnetic Resonance,
National High Magnetic Field Laboratory, Florida State University,
Tallahassee, Florida-32310, USA.}
\author{M. Pati and N. S. Dalal}
\affiliation{Department of Chemistry and Biochemistry, Florida State
University and NHMFL, Tallahassee, FL-32306, USA.}
\date{\today}

\begin{abstract}
Cr$^{5+}$ doped K$_3$NbO$_8$, considered to be useful as a electron
spin qubit, has been investigated by pulsed X-band ($\sim 9.7$ GHz)
and 240 GHz electron paramagnetic resonance and electron nuclear
double resonance (ENDOR). Comparison of the low temperature
electronic spin-lattice relaxation rate 1/$T_1$ at 9.7 and 240 GHz
shows that it is 250 times faster at 240 GHz than at X-band. On the
other hand, the spin-spin relaxation rate 1/$T_2$ appears largely
frequency independent and is very likely related to the
superhyperfine (SHF) coupling of the Cr$^{5+}$ electron with the
surrounding potassium and niobium nuclei. This coupling was
investigated by HYSCORE at 9.7 GHz and pulsed Mims ENDOR at 240 GHz.
The high frequency and field enabled us to unambiguously measure the
hyperfine and quadrupole couplings of the $^{39}$K in spite of its
small magnetic moment. We find that the largest $^{39}$K SHF
coupling is positive, with 0.522 MHz and 0.20 MHz as its isotropic
and dipolar parts respectively. $^{93}$Nb ENDOR was dominantly due
to its quadrapolar interaction, with a coupling of about 0.8 MHz,
and a SHF coupling of about 0.08 MHz. The significance of these data
to spin qubit studies is pointed out.
\end{abstract}

 \maketitle

\section{Introduction}
The Cr$^{5+}$ ion diluted in the diamagnetic host K$_3$NbO$_8$
(henceforth noted as Cr:K$_3$NbO$_8$) has been suggested as a
magnetic field calibration standard for high field electron
paramagnetic resonance (EPR) experiments~\cite{Cage99}. This $S =
1/2$ system exhibits a small linewidth ($\sim$ 0.15 - 0.2 mT
depending on the orientation) and the g-value just below the
free-electron g-value, g$_e$. The $I=0$ ($^{50,52,54}$Cr) isotopes
give rise to a single line that serves as a g-marker, while the
$^{53}$Cr (9.5\% $I=3/2$) yields four hyperfine lines that can be
used to calibrate the linearity of the field~\cite{Cage99}.

Recently Cr:K$_3$NbO$_8$ has also been suggested~\cite{Nellutla07}
as a new transition metal-ion based single electron-spin qubit
system for quantum information processing. Quantum computation
requires, among others, a long spin-lattice relaxation time $T_1$ as
well as a long spin-spin or spin-memory relaxation time $T_2$. The
latter time provides a measure of the decoherence processes in the
material and is used to calculate the figure of merit of a qubit,
which is defined as the number of operations that can be performed
before the phase information is lost. On the other hand, a 'reset'
of the qubit to a well defined starting state takes a time
proportional to $T_1$. Hence, shorter $T_1$ time means enhanced
computing speed. It is therefore desirable to understand the $T_1$
and $T_2$ relaxation processes via their temperature and frequency
dependence.

The present work reports on our detailed measurements of the
electronic $T_1$ and $T_2$ relaxation in Cr:K$_3$NbO$_8$ from
ambient temperatures down to 4 K at X-band ($\sim9.7$ GHz,
$\sim0.35$ tesla (T)) and at 240 GHz ($\sim8.7$ T). We find that the
low temperature relaxation time $T_1$ is strongly frequency
dependent: it decreases by a factor of 250 on going from 9.7 GHz to
240 GHz. On the other hand, $T_2$ was found to be frequency
independent in the investigated temperature range.

In order to understand the process(es) controlling the $T_2$ we
first employed hyperfine sublevel correlation (HYSCORE)
spectroscopy~\cite{Hoefer86} at X-band, and found that the nuclei
responsible for $T_2$ are the neighboring $^{39}$K nuclei. But the
magnitudes of $^{39}$K Zeeman, superhyperfine (SHF) and quadrupole
interactions are such that the analysis at X-band becomes very
complicated for any definitive analysis. On the other hand, the
utilization of the high frequency (240 GHz) pulsed EPR/ENDOR
spectrometer yielded ENDOR peaks that could easily be assigned to
the neighboring $^{39}$K as well as to the $^{93}$ Nb nuclei,
showing the advantage of the higher frequency/field EPR/ENDOR
spectroscopy. The data clearly suggests that the $T_2$ process, and
thus also the linewidths of the continuous wave (cw) EPR peaks, are
related to the unresolved SHF coupling to the $^{39}$K and $^{93}$Nb
nuclei. We note that there are few low frequency electron spin echo
envelope modulation~\cite{Barkhuijsen84} and
cw-ENDOR~\cite{DuVarney79, Dalal89, Dmitry08} studies where the
$^{39}$K hyperfine couplings were resolved.

This article is organized as follows. Section II provides the
experimental details on crystal structure, EPR and the ENDOR
spectrometers. The results and analysis are presented in section
III, and the salient points are summarized in section IV.

\section{Experimental details}

\begin{figure}
\includegraphics[width=7.25cm]{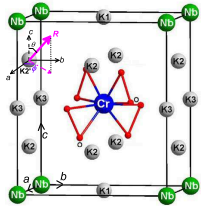}
\caption{\label{fXtal} (color online) Unit cell structure of
Cr:K$_3$NbO$_8$. Dodecahedral coordination geometry is shown only
for central ion for clarity. The potassium ions labeled as K2 and
K2$^\prime$ have an appreciable overlap with Cr 3$d_{x^{2}-y^{2}}$
orbital, whereas the ions labeled as K1 and K3 are in the unpaired
electron nodal plane. The definitions of the angles $\theta$ and
$\phi$ are shown by the vector $\emph{\textbf{R}}$ on K2$^\prime$.}
\label{fXtal}
\end{figure}

The Cr:K$_3$NbO$_8$ sample was synthesized according to the
published procedure~\cite{Cage99}. A typical synthesis involves
addition of CrO$_3$ to a solution of KOH and Nb$_2$O$_5$ at
-15$^{\circ}$C to form a 50\% ice slurry followed by drop-wise
addition of cold 30\% H$_2$O$_2$. Light yellow crystals of
Cr:K$_3$NbO$_8$ were obtained after keeping the final reaction
mixture at 5$^{\circ}$C for 2-3 days. The Cr$^{5+}$ concentration in
the studied Cr:K$_3$NbO$_8$ samples was determined to be $\sim 0.03$
mol\% by comparing its EPR signal intensity with a reference sample
(BDPA in polystyrene) of known number of spins.

The crystal structure of Cr$^{5+}$ doped K$_3$NbO$_8$ is shown in
Fig.~\ref{fXtal}. This system has body-centered tetragonal crystal
lattice ($I\bar{4}2m$ space group) with cell parameters $a = b =
6.7862(3)$ {\AA}, $c = 7.8312(7)$ {\AA} and $\alpha = \beta = \gamma
= 90{^\circ}$. The Cr$^{5+}$ ion occupies the Nb$^{5+}$ position and
for convenience is shown at the body-center position in
Fig.~\ref{fXtal}. Each Cr$^{5+}$/Nb$^{5+}$ ion is surrounded by four
peroxy (O$_2$$^{2-}$) groups (shown as red balls in
Fig.~\ref{fXtal}) in a dodecahedral arrangement.

The 9.7 GHz cw- and pulsed-EPR measurements were performed on a
commercial Bruker Elexsys 580 instrument with a Flexline dielectric
resonator using a single crystal of approximate size $2 \times 1
\times 1$ mm$^3$. The data at 240 GHz were collected on a home-built
superheterodyne high field cw-EPR instrument~\cite{Vantol05} which
is recently extended to pulsed operation~\cite{Morley08}. The single
crystals used were typically $0.2 \times 0.1 \times 0.1$ mm$^3$ in
size. A semi-confocal Fabry-Perot resonator was used with a 12.5 mm
spherical gold-coated mirror and a semi-transparent flat mirror,
consisting of a gold mesh deposited on a 160 $\mu$ thick quartz
cover slip. The maximum available power at this frequency is of the
order of 20 mW and the typical $\pi /2$ pulse length is 200 ns.

The ENDOR coil for the 240 GHz spectrometer consisted of a single
copper wire parallel to the mesh, which was led through a quartz
capillary (0.3 mm or 0.25 mm OD). The single crystal was attached to
the outside of the quartz capillary. A typical pulse sequence for
the Mims ENDOR experiment consisted of a stimulated echo sequence of
three $\pi/2$ pulses of about 240 ns length separated by 600 ns
($\tau$) and 200 $\mu$s ($t$) respectively. A 150 $\mu$s radio
frequency (RF) pulse was applied between second and third
millimeterwave pulses. About 10 W of power was applied to the
untuned ENDOR 'coil'. The pulsed ENDOR measurements were performed
at around 5-6 K. Typically 20 shots were averaged per RF frequency
and the total acquisition time of the ENDOR spectra was of the order
of 20-30 minutes.

\section{Results and Discussion}

\begin{figure}
\includegraphics[width=7.0cm]{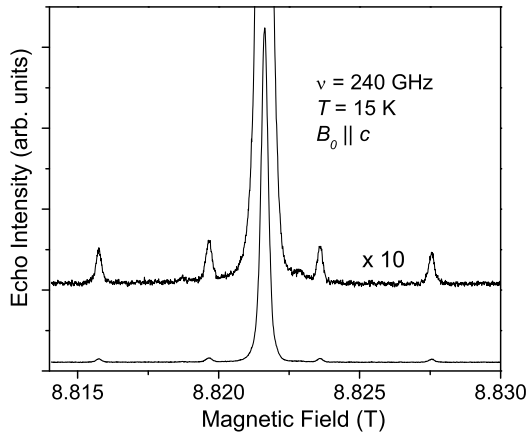}
\caption{\label{fEPR} Electron spin echo-detected EPR spectrum at
240 GHz and 15 K for $B_0$ $\parallel$ $c$-axis. The central peak is
from $I=0$ $^{50,52,54}$Cr nuclei whereas the four small peaks
separated by $\sim 4$ mT are from the hyperfine coupling of the
unpaired electron to $^{53}$Cr nucleus (9.5\% $I=3/2$).}
\end{figure}

Figure~\ref{fEPR} shows a typical 240 GHz electron-spin-echo (ESE)
detected EPR spectrum obtained with a two pulse Hahn echo sequence
($\pi/2$ - $\tau$ - $\pi$ - echo) and measuring the integrated echo
intensity. The ESE-EPR spectra of Cr:K$_3$NbO$_8$ show intense
central peak from the $I=0$ isotopes (4.34\% $^{50}$Cr, 83.8\%
$^{52}$Cr, 2.36\% $^{54}$Cr) at $g_\parallel = 1.9472 \pm 0.0002$
for the field $B_0 \parallel c$ and $g_\perp = 1.9878 \pm 0.0002$
for $B_0 \perp c$. The $g$-values indicate that the unpaired
electron resides in a 3$d_{x^{2}-y^{2}}$ orbital~\cite{Cage99,
Nellutla07, Dalal81}. The four peaks flanking the central peak are
the expected four nuclear transitions from the $I = 3/2$ of
$^{53}$Cr nucleus. The hyperfine coupling is $\sim 4$ mT for $B_0
\parallel c$ and $\sim 1$ mT for $B_0 \perp c$. All relaxation and
ENDOR measurements described below were performed on the central
$I=0$ peak.

\subsection{Spin-lattice relaxation}

\begin{figure}
\includegraphics[width=9.0cm]{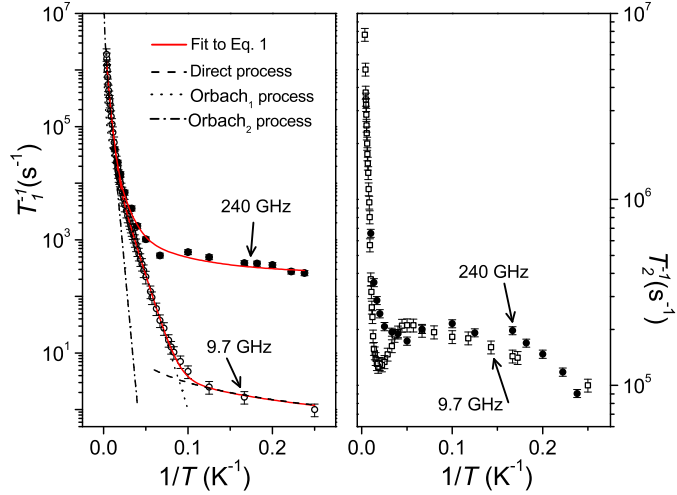}
\caption{\label{fT} (color online) Temperature dependence of 1/$T_1$
(left panel) and 1/$T_2$ (right panel) relaxation rates at 9.7 and
240 GHz for $B_0 \parallel c$-axis. The solid red lines in the left
panel are a fit to Eq. 1 with parameters listed in
Table~\ref{T1fitvalues}. The individual contributions to 1/$T_1$ are
also shown. See text for details.}
\end{figure}

The spin-lattice relaxation time $T_1$ of Cr:K$_3$NbO$_8$ was at 9.7
and 240 GHz measured by an inversion-recovery method employing the
sequences $\pi$ - $\tau$ - $\pi/2$ - FID and $\pi$ - t - $\pi/2$ -
$\tau$ - $\pi$ - $\tau$ - echo, respectively. The left panel in
Fig.~\ref{fT} shows the temperature dependence of 1/$T_1$ for $B_0
\parallel c$ at 9.7 and 240 GHz. At 9.7 GHz the $T_1$ increases
from about 500 ns at room temperature to about 1 s at 4 K. Above 25
K $T_1$ at 240 GHz is same as the 9.7 GHz but it reaches to $\sim4$
ms at 4 K. We attempted to fit the 9.7 and 240 GHz $T_1$ data to
different models~\cite{Abragam70} that included various combinations
of: (a) direct process with $T_1 ^{-1} \propto \coth(\hbar\omega
/2kT)$, (b) Raman process with $T_1 ^{-1} \propto T^{3+2m}$, where m
is the spectral dimensionality that depends on the spin system and
(c) Orbach process with $T_1 ^{-1} \propto 1/[e^{\Delta/kT}-1]$. The
best-fit to the data was found to be with the model given in Eq.(1).
\begin{equation}
T{_1}^{-1} = C{_1}\coth(\hbar\omega /2kT) +
C{_2}/[e^{\Delta{_1}/kT}-1] + C{_3}/[e^{\Delta{_2}/kT}-1]
\end{equation}
As can be seen from the solid lines in Fig.~\ref{fT}, the
temperature dependence of the spin-lattice relaxation rate 1/$T_1$
at 9.7 and 240 GHz was successfully modeled with Eq.(1) and the
best-fit parameters are listed in Table~\ref{T1fitvalues}.

\begin{table}
\caption{\label{T1fitvalues} Fit parameters of temperature
dependence of 1/$T_1$ at 9.7 GHz and 240 GHz. The values listed
correspond to the solid lines in Fig.~\ref{fT}.}
\begin{tabular} {|c|c|c|} \hline
Parameter & \multicolumn{2}{|c|}{Best-fit value} \\
\hline
& 9.7 GHz & 240 GHz \\
\hline
C$_1$ & 0.070(8) s$^{-1}$ & 250(8) s$^{-1}$ \\
C$_2$ & 6(1)*10$^{4}$ s$^{-1}$ & 6(1)*10$^{4}$ s$^{-1}$ \\
$\Delta_1$ & 75(3) cm$^{-1}$ & 75(3) cm$^{-1}$ \\
C$_3$ & 3.35(73)*10$^{6}$ s$^{-1}$ & 3.35(73)*10$^{6}$ s$^{-1}$ \\
$\Delta_2$ & 255(20) cm$^{-1}$ & 255(20) cm$^{-1}$ \\
\hline
\end{tabular}
\end{table}

It is noteworthy that the spin-lattice relaxation rate 1/$T_1$ at
low temperatures (4 - 10 K) is considerably faster at 240 GHz as
compared to that at 9.7 GHz which indicates that the direct process
is important at these temperatures. However, the ratio of the direct
process contribution to 1/$T_1$ at 240 and 9.7 GHz is not as large
as the expected $\omega ^5 \coth(\hbar\omega /2kT)$ dependence for a
Kramer's system~\cite{Abragam70} but it is closer to $\omega^3
\coth(\hbar\omega /2kT)$ dependence in non-Kramer's spin
systems~\cite{Abragam70}. For example, at 4 K the observed ratio of
the direct process contribution between 240 and 9.7 GHz is about 1/4
of the calculated $\omega^3 \coth(\hbar\omega /2kT)$ value whereas
it is about 4*10$^{-4}$ times the $\omega^5 \coth(\hbar\omega /2kT)$
value. This difference in the frequency dependence has been observed
before~\cite{Eaton01, Aminov97} and needs further theoretical
analysis which is beyond the scope of this undertaking.

As can be seen from Fig.~\ref{fT}, the spin-lattice relaxation rate
1/$T_1$ at 9.7 and 240 GHz frequencies for $T>$ 10 K is well
described by two Orbach relaxation pathways involving thermally
accessible energy levels at $\Delta_1 \sim 75$ cm$^{-1}$ and
$\Delta_2 \sim 255$ cm$^{-1}$. Based on the earlier infrared and
Raman studies~\cite{Haeuseler03}, the 75 cm$^{-1}$ mode can be
ascribed to the translational and librational modes, found to be
below 200 cm$^{-1}$ in the peroxychromate ion. Similarly, the 255
cm$^{-1}$ mode compares with the bending vibrational modes reported
in the 200 - 400 cm$^{-1}$ region in the peroxychromate ion.

\subsection{Spin-spin relaxation}

The standard 2-pulse Hahn echo sequence was used to obtain the
spin-spin relaxation time $T_2$ at both 9.7 and 240 GHz. The right
panel of Fig.~\ref{fT} shows the 1/$T_2$ rate as a function of
temperature for $B_0 \parallel c$. At 9.7 GHz the $T_2$ increases
from $\sim 150$ ns at room temperature to $\sim8$ $\mu$s at 50 K,
then decreases to $\sim5$ $\mu$s at 20 K and finally increases to
$\sim10$ $\mu$s at 4 K. In contrast to $T_1$, the $T_2$ time does
not show a strong field dependence. However, the temperature
dependence of $T_2$ at both frequencies shows a minimum at about 20
K. We tentatively ascribe this minimum to some sort of motional
narrowing caused by the increase in the spin fluctuations of K and
Nb hyperfine fields as the temperature is raised above about 20 K.
This type of temperature dependence has previously been reported for
dilute solutions of free radicals~\cite{Brown74, Brown76,
Stillman80} and other Cr$^{5+}$ complexes~\cite{Nakagawa92}. Above
50 K, the spin-spin relaxation time $T_2$ is mainly governed by the
spin-lattice relaxation processes.

\subsection{Superhyperfine (SHF) interactions}

As mentioned earlier the hyperfine interaction of the unpaired
electron with the $^{53}$Cr in Cr:K$_3$NbO$_8$ was explored
previously~\cite{Dalal81}, and therefore in this report we will only
address the SHF coupling with the potassium ($I = 3/2$ $^{39}$K) and
niobium ($I = 9/2$ $^{93}$Nb) nuclei. We believe that the electronic
spin-spin relaxation processes in Cr:K$_3$NbO$_8$ are most likely
influenced by these interactions and in order to characterize them
we used HYSCORE spectroscopy at 9.7 GHz and pulsed Mims ENDOR at 240
GHz.

\begin{figure}
\includegraphics[scale=0.90]{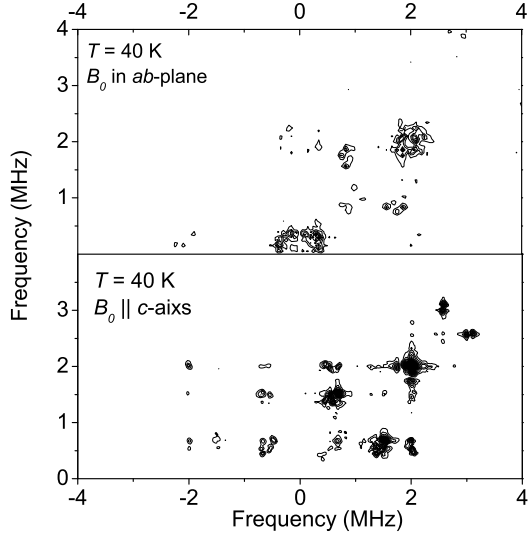}
\caption{\label{fHyscore} HYSCORE spectra of Cr:K$_3$NbO$_8$ at 9.7
GHz and 40 K. The top panel represents $B_0$ in the $ab$-plane and
the bottom panel for $B_0 \parallel c$-axis.}
\end{figure}

The HYSCORE spectra of Cr:K$_3$NbO$_8$ at 40 K are shown in
Fig.~\ref{fHyscore} for the orientation along the $c$-axis (bottom
panel) and an orientation in the $ab$-plane (top panel). The spin
echo modulations due to the interactions with neighboring potassium
nuclei are very large, which resulted in a good quality HYSCORE
spectrum. However, the spectra were found to be too complex for
analysis, possibly because at 9.7 GHz, the Zeeman, quadrupole and
SHF interactions become comparable and lead to a complex interplay
of these interactions. We therefore resorted to ENDOR measurements
at 240 GHz ($\sim8.7$ T) where the Zeeman terms become dominant.

\begin{figure}
\includegraphics[scale = 1.0]{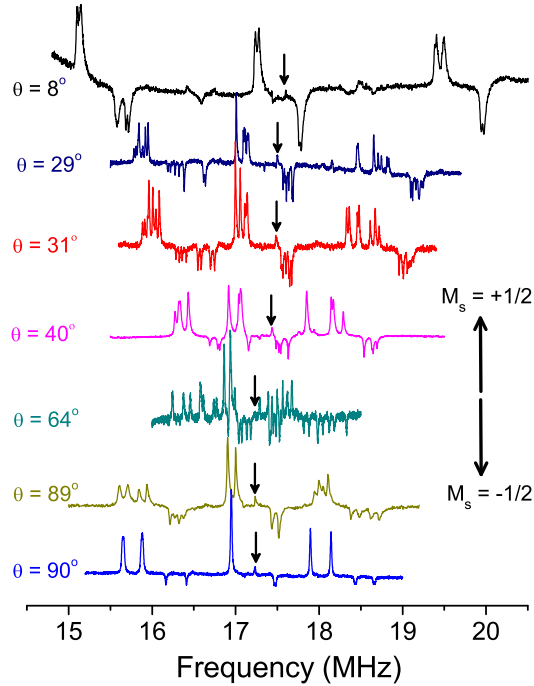}
\caption{\label{fMims} (color online) Mims ENDOR spectra of $^{39}$K
at 240 GHz and 5 K at several crystal orientations. $\theta$
indicates the angle between $B_0$ and the $c$-axis. Echo intensity
increases during the ENDOR transitions in the $M_s$ = +1/2 manifold
(upward peaks), while it decreases for the ENDOR transitions in the
$M_s$ = -1/2 manifold (downward peaks). The arrows correspond to the
nuclear Zeeman frequency of $\sim17.25$ MHz.}
\end{figure}

At 240 GHz, pulsed ENDOR signals of $^{39}$K were measured at
various crystal orientations (see Fig.~\ref{fMims}). A systematic
study of orientation dependence of the ENDOR could not be performed
due to the unavailability of a goniometer for this spectrometer at
the present time. While the angle ($\theta$) of $B_0$ relative to
the crystal $c$-axis can be accurately determined from the resonance
field of the EPR transition (via the g-tensor anisotropy), the axial
g-tensor symmetry does not help with the precise determination of
the angle $\phi$ of $B_0$ in the $ab$-plane. However, a closer look
at the crystal structure (see Fig.~\ref{fXtal}) reveals that only
the eight potassium nuclei labeled as K2 and K2$^\prime$ have an
appreciable overlap with the Cr$^{5+}$ 3$d_{x^2-y^2}$ orbital and
therefore are expected to be responsible for the $^{39}$K ENDOR
signals. These eight equivalent nuclei are related by symmetry
($xyz, \bar{x}yz, x\bar{y}z, \bar{x}\bar{y}z, yx\bar{z},
\bar{y}x\bar{z}, y\bar{x}\bar{z}, \bar{y}\bar{x} \bar{z}$), and at
some orientations eight sets of signals can indeed be distinguished
(e.g. $\theta = 31^\circ$ spectrum in Fig.~\ref{fMims}). The
potassium ions labeled as K1 and K3 are in the nodal planes of the
Cr unpaired electron. They should thus exhibit negligible isotropic
SHF couplings.

It should be noted that both negative (smaller stimulated echo) and
positive (larger stimulated echo) signals are observed, and that the
signal intensities are quite large. When on ENDOR resonance, the
observed echo height varied from 40\% to 200\% of the echo height
observed without RF power or out of resonance. The ENDOR intensity
is largest when the shot-repetition time is of the order of $T_1$.
This implies that the $^{39}$K nuclei are strongly polarized by the
pulse sequence, leading to an increase in echo intensity when a
nuclear transition in the $M_s = +1/2$ electron manifold is
addressed, while the echo intensity becomes smaller when a
transition in the $M_s = -1/2$ state is excited, as explained in
Ref.~\cite{Bennebroek97}. These anomalous ENDOR intensities occur
when the microwave quantum ($\hbar\omega$) is of the same order or
larger than the thermal energy $kT$ and allows for the determination
of the sign of the hyperfine coupling in both
pulsed~\cite{Bennebroek97, Epel01} and cw-ENDOR~\cite{Vantol99}. The
observed sets of $^{39}$K ENDOR signals are sufficient for a unique
assignment of the SHF and quadrupole tensor values and their
relative orientations. The data are analyzed by the diagonalization
of the usual effective spin Hamiltonian given in Eq.(2) with $S=1/2$
and $I=3/2$.
\begin{equation}
\hat{H} = \mu_B \vec{B_0}\tilde{g}\hat{S} +
\sum_n{(\hat{S}\tilde{A_n}\hat{I_n} - \gamma_n\vec{B_0}\hat{I_n} +
\hat{I_n}\tilde{Q_n}\hat{I_n})}
\end{equation}
Here, $\tilde{g}$ is the electron Zeeman tensor, $\tilde{A}$ and
$\tilde{Q}$ are, respectively, SHF and quadrupole tensors of the
surrounding nuclei. The principal values and their directions of the
$\tilde{A}$ tensor (namely A$_1$, A$_2$, A$_3$) and $\tilde{Q}$
tensor (namely Q$_1$, Q$_2$, Q$_3$) for K2$^\prime$ nucleus obtained
from the analysis are given in Table~\ref{t_tensors}. The tensors
for the other chemically equivalent K2 nuclei can be obtained via
the afore-mentioned crystal symmetry operations. The isotropic SHF
interaction is +0.522 MHz, and can be interpreted as a spin density
of 0.23\% in the $4s$ orbital when compared to the hyperfine
coupling in the ground state of atomic potassium~\cite{Li94}. While
we cannot make a completely unambiguous assignment of the specific
potassium nucleus associated with each of the eight sets of SHF and
quadrupole tensors, it is natural to assume that the anisotropic
contribution to the SHF interaction should reflect the dipolar
interaction between the potassium nuclear spin and the chromium
electron spin. We therefore assign the SHF tensor given in
Table~\ref{t_tensors} to the potassium for which the $A_3$ direction
is close to the Cr-K direction (labeled as K2$^\prime$ in the
Fig.~\ref{fXtal}).

\begin{table}
\caption{\label{t_tensors} The principal values and their directions
of $^{39}$K SHF and quadrupole tensors determined by 240 GHz Mims
ENDOR. These values correspond to K2$^\prime$ and the values for the
remaining seven K2 potassium ions can be obtained from the symmetry
relations given in the text. $\theta$ is the angle with the $c$-axis
and $\phi$ the angle in the $ab$ plane (see Fig.~\ref{fXtal}). The
errors in the directions are estimated to be 2 degrees.}
\begin{tabular} {|c|c|c|c|} \hline
 & Value(MHz) & $\theta$ & $\phi$ \\
\hline
A$_1$ & +0.407(5) & 147 & 90 \\
A$_2$ & +0.425(5) & 90 & 0 \\
A$_3$ & +0.733(5) & 57 & 90 \\
Q$_1$ & -0.423(4) & 87.6 & 65.6 \\
Q$_2$ & -0.329(4) & 90.6 & -24 \\
Q$_3$ & 0.752(4)  & 2.5 & -100 \\
\hline
\end{tabular}
\end{table}

The quadrupole tensor of K2$^\prime$ is not entirely axial and has
the direction of its largest principal value close to the $c$-axis.
The slight deviation from the $c$-axis indicates that the local
surrounding of the Cr$^{5+}$ ion is slightly distorted as compared
to that for Nb$^{5+}$ ion in the undoped lattice.

\begin{figure}
\includegraphics[scale = 0.90]{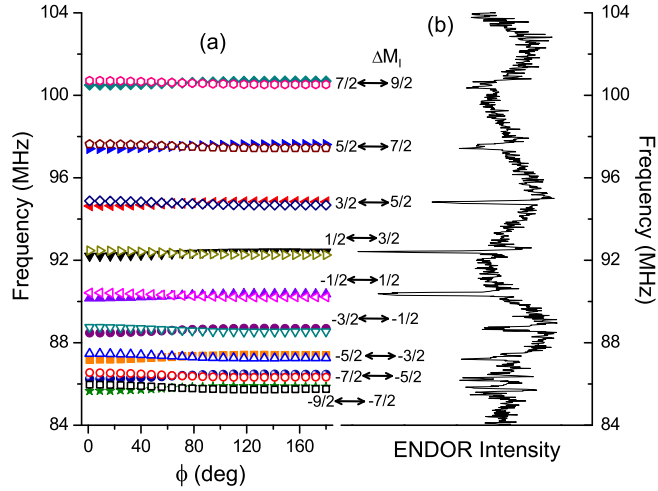}
\caption{\label{fNbENDOR} (color online) (a) Calculated energy
levels as a function of $\phi$ at 240 GHz for $S = 1/2$ and $I =
9/2$ for $B_0$ 40$^\circ$ from the $c$-axis. Closed and open symbols
shown for each $\Delta M_I$ transition correspond to $M_s$ = +1/2
and -1/2. (b) Experimental $^{93}$Nb Mims ENDOR at 5 K and 240 GHz
for the same orientation.}
\end{figure}

At some crystal orientations $^{93}$Nb $(I=9/2)$ ENDOR is also
observed. ENDOR spectra for field about 40$^\circ$ from the crystal
$c$-axis is shown in Fig.~\ref{fNbENDOR}(b). We speculate that at
most orientations the SHF interaction is too small to give an
appreciable ENDOR intensity, but it can be detected when the field
is along the Nb-Cr dipolar direction because of the largest dipolar
interaction in this direction. Since Cr and Nb ions are fairly well
separated (Cr-Nb: $\sim6.2$ \AA), the dipolar contribution to the
Cr-Nb SHF interaction can be estimated using a point dipolar
approximation as b[3$\cos^{2}(\psi)$-1] with b $\sim 0.08$ MHz,
where $\psi$ is the angle between Cr-Nb dipolar axis and $B_0$.
While the limited data prevents a detailed analysis, the $^{93}$Nb
ENDOR spectrum in Fig.~\ref{fNbENDOR} can be simulated with a
quadrupole tensor Q[3$\cos^{2}(\theta)$-1] with $\mid$Q$\mid$ = 0.82
MHz under the assumption that the quadrupole tensor is nearly axial
with the principal direction along the crystal $c$-axis.

\section{Summary}
We have measured the electron spin relaxation times in
Cr:K$_3$NbO$_8$ using pulsed EPR at 9.7 and 240 GHz. The SHF
couplings of the electron spin with the neighboring potassium nuclei
have also been determined using pulsed Mims ENDOR at 240 GHz. The
ENDOR results show that the SHF interaction of the Cr unpaired
electron with neighboring potassium (labeled as K2 and K2$^\prime$
in Fig.~\ref{fXtal}) and the niobium nuclei is responsible for the
electron spin-spin relaxation $T_2$ process. An unpaired spin
density of about 0.2\% on the K2 and K2$^\prime$ nuclei has been
estimated from the isotropic $^{39}$K SHF coupling constant. While
the $T_2$ time in Cr:K$_3$NbO$_8$ is largely frequency independent,
its temperature dependence shows a minimum around 20 K that has been
tentatively ascribed to the change in the rate of surrounding
nuclear spin flip flops. At both 9.7 and 240 GHz the spin-lattice
relaxation time $T_1$ is dominated by a direct and two Orbach
processes. The modes involved in the Orbach processes are assigned
to bending vibrations, translational lattice vibrations and/or
libration modes of the CrO$_8^{3-}$ ion. Furthermore, the direct
process contribution to $T_1$ at 240 GHz is found to be about more
than two orders of magnitude larger than the contribution at 9.7
GHz. Surprisingly, the contribution ratio is closer to an $\omega ^3
\coth(\hbar\omega /2kT)$ dependence in a non-Kramer's system than it
is to the $\omega ^5 \coth(\hbar\omega /2kT)$ dependence expected
for a Kramer's system. Further theoretical analysis is necessary to
understand this divergence. Finally, the frequency dependence of
relaxation times in Cr:K$_3$NbO$_8$ shows that it is possible to
tune $T_1$ with the frequency/field while keeping $T_2$ relatively
the same, a feature useful in transition metal-ion based spin qubits
for quantum information applications.

\begin{acknowledgments}
We thank Prof. Ronald J. Clark of Chemistry and Biochemistry
department of the Florida State University for crystal structure
determination of K$3$NbO$_8$. SN and JvT acknowledge the State of
Florida and NSF Cooperative Agreement grant DMR0654118 and NSF grant
DMR0520481 for financial support. NSD acknowledges NSF for funding
through grant DMR0506946.
\end{acknowledgments}


\begin{thebibliography}{mm}

\bibitem{Cage99}
B. Cage, A. Weekley, L. -C. Brunel and N. S. Dalal, Anal. Chem.
\textbf{71}, 1951 (1999).

\bibitem{Nellutla07}
S. Nellutla, K.-Y. Choi, M. Pati, J. van Tol, I. Chiorescu and N. S.
Dalal, Phys. Rev. Lett. \textbf{99}, 137601 (2007).

\bibitem{Hoefer86}
P. Hoefer, A. Grupp, G. Nebenfuehr and M. Mehring, Chem. Phys. Lett.
\textbf{132}, 279 (1986).

\bibitem{Barkhuijsen84}
H. Barkhwjsen, R. de Beer, A. F. Deutz, D. van Ormondt and G.
V\"{o}lkel, Solid State Comm. \textbf{49}, 679 (1984).

\bibitem{DuVarney79}
R. C. DuVarney and J. M. Spaeth, Solid State Comm. \textbf{32}, 1237
(1979).

\bibitem{Dalal89}
N. S. Dalal and P. K. Kahol, Solid State Comm. \textbf{70}, 623
(1989).

\bibitem{Dmitry08}
D. Zverev, H. Vrielinck, F. Callens, P. Matthys, S. Van Doorslaer
and N. M. Khaidukov, Phys. Chem. Chem. Phys. \textbf{10}, 1789
(2008).

\bibitem{Vantol05}
J. van Tol, L. C. Brunel and R. J. Wylde, Rev. Sci. Inst.
\textbf{76}, 074101 (2005).

\bibitem{Morley08}
G. M. Morley, L.C. Brunel and J. van Tol, Rev. Sci. Inst. (to be
published). Article is avaliable at http://arxiv.org/abs/0803.3054.

\bibitem{Dalal81}
N. S. Dalal, J. M. Millar, M. S. Jagadeesh and M. S. Seehra, J.
Chem. Phys. \textbf{74}, 1916 (1981).

\bibitem{Schweiger01}
A. Schweiger and G. Jeschke, \emph{Principles of Pulse Electron
Paramagnetic Resonance} (Oxford University Press, Oxford, 2001).

\bibitem{Abragam70}
A. Abragam and B. Bleaney, \emph{Electron Paramagnetic Resonance of
Transition ions} (Clarendon Press, Oxford, 1970).

\bibitem{Eaton01}
S. S. Eaton, J. Harbridge, G. A. Rinard, G. R. Eaton and R. T.
Weber, Appl. Magn. Reson. \textbf{20}, 151 (2001).

\bibitem{Aminov97}
L. K. Aminov, I. N. Kurkin, S. P. Kurzin, D. A. Lukoyanov, I. Kh.
Salikhov and R. M. Rakhmatullin, JETP \textbf{84}, 183 (1997).

\bibitem{Haeuseler03}
H. Haeuseler and G. Haxhillazi, J. Raman Spectrosc. \textbf{34}, 339
(2003).

\bibitem{Brown74}
I. M. Brown, J. Chem. Phys. \textbf{60}, 4930 (1974).

\bibitem{Brown76}
I. M. Brown, J. Chem. Phys. \textbf{65}, 630 (1976).

\bibitem{Stillman80}
A. E. Stillman, L. J. Schwartz and J. H. Freed, J. Chem. Phys.
\textbf{73}, 3502 (1980).

\bibitem{Nakagawa92}
K. Nakagawa, M. B. Candelaria, W. W. C. Chik, S. S. Eaton and G. R.
Eaton, J. Magn. Reson. \textbf{98},81 (1992) and the references
therein.

\bibitem{Bennebroek97}
M. T. Bennebroek and J. Schmidt, J. Magn. Reson. \textbf{128}, 199
(1997).

\bibitem{Epel01}
B. Epel, A. P\"{o}ppl, P. Manikandan, S. Vega and D. Goldfarb, J.
Magn. Reson. \textbf{148}, 388 (2001).

\bibitem{Vantol99}
J. van Tol, L. C. Brunel and P. Wyder, Bull. Am. Phys. Soc.
\textbf{44}, 1130 (1999).

\bibitem{Li94}
S. Li and J. F. Clauser, Phys. Rev. A. \textbf{49}, 2702 (1994).

\end{thebibliography}
\end{document}